\documentclass[onecolumn,amsmath,nofootinbib,secnumarabic,eqsecnum,amssymb,12pt]{revtex4}

\usepackage{bm}

\usepackage{epsf}
\usepackage{graphicx,epsfig}
\usepackage{amsfonts}
\usepackage{amssymb}

\usepackage[pdftex]{hyperref}

\def\t{\tau}
\def\e{\epsilon}
\def\eps{\epsilon}
\def\c{c_s}
\def\cs{c_s}
\def\et{\eta}
\def\k{k}
\def\Tdot#1{{{#1}^{\hbox{.}}}}

\def\bk{\mathbf k}
\def\bx{\mathbf x}
\def\CO{{\cal O}}

\def \invc{\left( \frac{1}{\c^2}-1\right)}
\def\l{\ln \frac{\a}{2}}
\def\half{\frac12}

\def\PX{P_{,X}}

\def\bkone{\mathbf k_1}
\def\bktwo{\mathbf k_2}
\def\bkthree{\mathbf k_3}
\def\bqone{\mathbf q_1}
\def\bqtwo{\mathbf q_2}

\def\zl{\zeta_l}
\def\zs{\zeta_s}
\def\zn{\zeta_n}
\def\tb{\bar t}

\def\F{F}
\def\Pz{P_{\zeta}}

\def\a{\alpha_K}
\def\bc{{ C}}

\def\fNL{f_{NL}}

\newcommand{\bea}{\begin{eqnarray}}
\newcommand{\eea}{\end{eqnarray}}
\newcommand\be{\begin{equation}}
\newcommand\ee{\end{equation}}

\newcommand{\nn}{\nonumber}

\newcommand{\refeq}[1]{(\ref{#1})}

\makeatletter
\renewcommand\section{\@startsection {section}{1}{\z@}%
                                 {-3.5ex \@plus -1ex \@minus -.2ex}
                                   {2.3ex \@plus.2ex}%
                                   {\normalfont\large\bfseries}}
\renewcommand\subsection{\@startsection{subsection}{2}{\z@}%
                                   {-3.25ex\@plus -1ex \@minus -.2ex}%
                                     {1.5ex \@plus .2ex}%
                                     {\normalfont\bfseries}}
\renewcommand\subsubsection{\@startsection{subsubsection}{3}{\z@}%
                                   {-3.25ex\@plus -1ex \@minus -.2ex}%
                                     {1.5ex \@plus .2ex}%
                                     {\normalfont\bf\itshape}}
                                    
\makeatother

\begin{document}

\begin{center}
{\large \bf On the squeezed limit of the bispectrum in  \\
\vspace{0.5cm}
general single field inflation}

\vspace*{0.5in} {S\'ebastien Renaux-Petel\footnote{renaux@apc.univ-paris7.fr}  }
\\[.3in]
{\em   APC (Astroparticules et Cosmologie) \\
UMR 7164 (CNRS, Universit\'e Paris 7, CEA, Observatoire de
Paris)\\
10, rue Alice Domon et L\'eonie Duquet,
 75205 Paris Cedex 13, France  \\[0.3in]}
\end{center}

\begin{center}
{\bf
Abstract}
\end{center}

We investigate the consistency relation relating the squeezed limit of the bispectrum to the scalar spectral index in single field models of inflation. We give a simple integral formula for the bispectrum in the squeezed limit in terms of the free mode mode functions of the primordial curvature perturbation, in any Lorentz invariant single field model of inflation and without resorting to any approximation, generalizing a recent result obtained by Ganc and Komatsu in the case of canonical kinetic terms. We use our result to verify the consistency relation in an exactly solvable class of models with a non-trivial speed of sound. We then verify the consistency relation at the first non-trivial order in the slow-varying approximation in general single field inflation (a known result) and at second order in this approximation in canonical single field inflation.

\noindent

\vfill

\newpage
\setcounter{page}{1}

\tableofcontents

\newpage

\section{I\lowercase{ntroduction}}

Observations of the anisotropies of the Cosmic Microwave Background radiation reveal that the primordial fluctuations which seeded them are to a good approximation adiabatic --- in which case they are solely characterized by a primordial curvature perturbation $\zeta$ --- scale-invariant and Gaussian \cite{Komatsu:2010fb}. However, a small amount of non-Gaussianity is still allowed by the data and the information contained in this non-Gaussian component will contribute to huge advance in our understanding of the early universe by discriminating between otherwise competing models (see for instance \cite{Chen:2010xk,Koyama:2010xj} for recent reviews and \cite{Liguori:2010hx,Komatsu:2010hc,Yadav:2010fz} for more observational aspects). One of the most important realizations in this respect is the identification of a consistency relation between the primordial two-point correlation function --- the power spectrum --- and a particular geometrical limit of the three-point correlation function --- the bispectrum --- that is valid in any single field model of inflation \cite{malda,Creminelli:2004yq}:
\be
\lim_{k_3 \to 0} \langle \zeta_{\bkone} \zeta_{\bktwo}\zeta_{\bkthree} \rangle = -(2 \pi)^3  \delta^{(3)} (\sum_i \bk_i)(n_s(k_1)-1)P_{\zeta}(k_1)P_{\zeta}(k_3) 
\label{consistency}
\ee
where
\be
\langle \zeta_{\bkone} \zeta_{\bktwo}\rangle = (2 \pi)^3  \delta^{(3)} (\bkone + \bktwo)P_{\zeta}(k_1) 
\ee
and 
\be
n_s(k)-1=  \frac{ d \, {\rm ln }\left[ k^3P_{\zeta}(k)\right]}{ d\, {\rm ln} k}
\ee
is the scalar spectral index. Originally derived by Maldacena in his study of the bispectrum generated by a phase of slow-roll single field inflation \cite{malda}, the relation \refeq{consistency} has later been generalized by Creminelli and Zaldarriaga to any single field model upon using very general arguments \cite{Creminelli:2004yq}. To understand its theoretical and observational relevance, let us remind that the observational constraints on the primordial bispectrum are most often quoted in terms of the dimensionless momentum-dependent function $\fNL(k_1,k_2,k_3)$ defined by
\be
 \langle \zeta_{\bkone} \zeta_{\bktwo}\zeta_{\bkthree} \rangle = (2 \pi)^3  \delta^{(3)} (\sum_i \bk_i) \frac{6}{5} \fNL  \left( P_{\zeta}(k_1)P_{\zeta}(k_3)+{\rm 2\, permutations}  \right)
 \label{def-fnl}
\ee
Comparing \refeq{consistency} and \refeq{def-fnl}, and using the fact that $\Pz(k) \propto k^{-3+(n_s-1)}$, one finds that
\be
\fNL^{sq}(k_1)=\frac{5}{12} (1-n_s(k_1))
\label{theo-2}
\ee
where we have defined 
\be
\fNL^{sq}(k_1)= \lim_{k_3 \to 0}  \fNL(k_1,k_2,k_3)\,
\ee
and where ${}^{sq}$ qualifies the ``squeezed'' configuration of momenta under consideration $(k_3 \to 0, k_1 \simeq k_2)$. Given that the deviation of the primordial power spectrum from scale invariance is tightly constrained, $n_s=0.963 \pm 0.012\, (68 \% CL)$ \cite{Komatsu:2010fb}, any convincing detection of a large bispectrum signal in the squeezed limit ($\fNL^{sq} \gtrsim 1$) would hence rule out all models of inflation based on a single scalar field.

Because of this far-reaching implication, it is very important to understand every aspect of the above consistency relation and to verify it by explicit calculations of the primordial bispectrum. This has been done for instance in references \cite{Seery:2005wm,Chen:2006nt,Cheung:2007sv} at first order in a slow-varying approximation. Recently however, Ganc and Komatsu gave an integral formula for the squeezed bispectrum in single field inflationary models with canonical kinetic terms that is valid without any approximation \cite{Ganc:2010ff} . Although they were not able to derive the consistency relation from it, they used their approach to verify the relation \refeq{consistency} non-perturbatively in an exactly solvable class of models.

In this paper, we extend their work in several directions. After reviewing certain aspects of the most general Lorentz-invariant single field models of inflation in section \ref{General}, we generalize the main result of Ganc and Komatsu by giving an integral formula for the squeezed bispectrum in this general class of models without resorting to any approximation (section \ref{general-formula}). We then use this result to verify Maldacena's consistency relation in specific cases in section \ref{verification}. We first consider an exactly solvable class of models with a non-trivial speed of sound. Second, we verify the consistency relation in general single field inflation at first order in the slow-varying approximation. Although this result has already been obtained in references \cite{Chen:2006nt,Cheung:2007sv}, we believe our different derivation to be useful. We demonstrate this by finally deriving the consistency relation in canonical single field inflation at \textit{second order} in the slow-varying approximation. We give our conclusions in section \ref{conclusion} and leave the details of some long calculations to the appendices.

\section{A \lowercase{reminder of general single field inflation}}
\label{General}

In this section, we wet up our notation and we give the second- and third-order scalar action in general single-field inflation that were calculated in \cite{Garriga:1999vw,Chen:2006nt} and that we will use in the following. 

The action we consider takes the form (we use units in which $\hbar=c=M_{pl}=1$) \cite{Garriga:1999vw}
\begin{equation} \label{general}
S=\int d^4x \sqrt{-g} \left[\frac12 R +
P(X,\phi)\right]
\end{equation}
where $\phi$ is the inflaton, $X\equiv -\frac{1}{2}g^{\mu\nu}\partial_{\mu}\phi \partial_{\nu}\phi$ and $P(X,\phi)$ is the most general Lorentz-invariant Lagrangian that is a function of $\phi$ and of its first derivative (see \cite{Langlois2008,Gao:2008dt,Langlois:2008qf,Arroja:2008yy} for extensions to multifield inflationary models). In a spatially flat Friedmann-Lema\^itre-Robertson-Walker spacetime, with metric
\be
ds^2=-dt^2+a^2(t) d{\bx}^2\,,
\ee
where $t$ is cosmic time, the scalar field $\phi$ is homogenous and its energy-momentum tensor reduces to that of a perfect fluid with energy density
\be
\rho=2X \PX-P
\ee
and pressure $P$. The dynamics of the scale factor and of the inflation field are then governed by the Friedmann equations
\bea
3 H^2&=&\rho \\
\dot \rho&=&-3H(\rho+P)
\label{continuity}
\eea
(the equation of motion of the scalar field reducing to the continuity equation \refeq{continuity}). 

In the following, we assume that the scalar field Lagrangian $P(X,\phi)$ is such that a prolonged stage of inflation occurs and we study the cosmological perturbations generated in such scenarios. In this respect, it is useful to introduce the parameters
\bea
\c^2&\equiv&\frac{\PX}{\PX+2X P_{,XX}}  \,, \label{cs} \\
\eps &\equiv& -\frac{\dot H}{H^2}=\frac{\dot \phi^2 \PX}{2 H^2}\,, \qquad \eta \equiv \frac{\dot \eps}{H \eps} \,, \qquad s \equiv \frac{\dot \cs}{H \cs}\,, \label{eps}\\
\Sigma&=&X P_{,X}+2X^2P_{,XX}  = \frac{H^2\epsilon}{c_s^2} ~,\\
\lambda&=& X^2P_{,XX}+\frac{2}{3}X^3P_{,XXX} 
\label{lambda}
\eea
where $\cs^2$ is known as the ``speed of sound'' squared of perturbations, that we require to be comprise between $0$ and $1$ to avoid any pathological behaviour. When the scalar field Lagrangian is canonical, $P=X-V(\phi)$, the speed of sound equals one and the parameters $s$ and $\lambda$ identically vanish.

To compute the action at second and cubic order in the perturbations, it is useful to work in the ADM formalism \cite{adm} in which the metric is written in the form
\be
ds^2=-N^2 d t^2+h_{ij} (dx^i+N^i  dt)(d x^j+N^j dt)\,.
\ee
The lapse $N$ and the shift $N^i$ appear indeed as Lagrange multipliers in the action \refeq{general} and hence can be algebraically  expressed in terms of the true physical degrees of freedom. Restricting to scalar perturbations, there is only one such quantity, namely, the gauge invariant scalar perturbation $\zeta$ that is constant outside the horizon and of which we want to determine the statistical properties. $\zeta$ is most easily defined in the comoving gauge where the inflaton $\phi$ is homogeneous and where the three-dimensional metric $h_{ij}$ takes the form
\be
h_{ij}=a^2 e^{2 \zeta} \delta_{ij}\,.
\ee
Solving the constraint equations for $N$ and $N^i$ in terms of $\zeta$ and plugging them back into the action \refeq{general}, one then finds the second-order action \cite{Garriga:1999vw}
\begin{eqnarray}
S_2 = \int dt d^3x~
\left[a^3  \frac{\epsilon}{c_s^2}\dot\zeta^2- a \epsilon
(\partial \zeta)^2 \right] 
\label{actionQuad}
\end{eqnarray}
and third-order action \cite{Chen:2006nt}
\begin{eqnarray}
S_3&=&\int dt d^3x \left[
-a^3 (\Sigma(1-\frac{1}{c_s^2})+2\lambda)\frac{\dot{\zeta}^3}{H^3}
+\frac{a^3\epsilon}{c_s^4}(\epsilon-3+3c_s^2 )\zeta\dot{\zeta}^2
\right.
\cr
&&
\left.
+\frac{a\epsilon}{c_s^2}(\epsilon-2s+1-c_s^2)\zeta(\partial\zeta)^2-
a \frac{\epsilon (4-\epsilon)}{2c_s^2}\dot{\zeta}(\partial
\zeta)(\partial \chi)+\frac{a^3\epsilon}{2c_s^2}    \Tdot{\left(\frac{\et}{\c^2}\right)} \zeta^2\dot{\zeta} \right.
\cr
&&
\left.
+\frac{\eps}{4a} (\partial^2 \zeta) \chi_{,i}\chi^{,i}
- f(\zeta)\frac{\delta L}{\delta \zeta}|_1 \right]
 \label{action3}
\end{eqnarray}
where
\be
\chi \equiv \partial^{-2} \left( a^2\frac{\eps}{\cs^2} {\dot \zeta} \right),
\ee
$\partial^{-2}$ denotes the inverse Laplacian and, in the last term
\begin{eqnarray}
\frac{\delta
L}{\delta\zeta}\mid_1 = \frac{\delta S_2}{\delta\zeta}  = - \frac{\partial}{\partial t} \left(2a^3\frac{\eps}{\cs^2}\dot \zeta\right) +2a \eps \partial^2 \zeta
\label{variation-S2}
\end{eqnarray}
and
\begin{eqnarray} \label{redefinition}
f(\zeta)&=&\frac{\eta}{4c_s^2}\zeta^2+\frac{1}{c_s^2H}\zeta\dot{\zeta}+
\frac{1}{4a^2H^2}[-(\partial\zeta)(\partial\zeta)+\partial^{-2}(\partial_i\partial_j(\partial_i\zeta\partial_j\zeta))] \nonumber \\
&+&
\frac{1}{2a^2H}[(\partial\zeta)(\partial\chi)-\partial^{-2}(\partial_i\partial_j(\partial_i\zeta\partial_j\chi))] ~.
\end{eqnarray}

\section{T\lowercase{he squeezed limit of the bispectrum in general single field inflation}}
\label{general-formula}

In this section, we give an explicit integral form for the bispectrum generated during a phase of general single field inflation in the squeezed limit, generalizing the result of \cite{Ganc:2010ff}. We closely follow their derivation, to which we refer the reader for more details. Let us nonetheless now outline the strategy and give important precisions before mowing to the calculation itself.\\

Our final goal is to compute the bispectrum $ \langle \zeta_{\bkone}(\tb) \zeta_{\bktwo}(\tb) \zeta_{\bkthree}(\tb) \rangle$ in the extreme squeezed limit in which $k_3 \to 0$ (and hence  $k_1\simeq k_2$) at a time $\tb$ where all the three $\zeta_{\bk_i}$ have become classical and have reached their constant value, typically a few-efolds after the largest momenta have crossed the sound horizon at $k_{1,2} \cs \approx a H$. We will first calculate $ \langle \zeta_{\bkone} \zeta_{\bktwo} \rangle_{\bkthree}$ (the expectation value of $ \zeta_{\bkone} \zeta_{\bktwo}$ given that $\zeta_{\bkthree}$ has a particular value) and then correlate this result with $\zeta_{\bkthree}$ to find $ \langle \zeta_{\bkone} \zeta_{\bktwo} \zeta_{\bkthree} \rangle= \langle  \langle \zeta_{\bkone} \zeta_{\bktwo} \rangle_{\bkthree} \zeta_{\bkthree} \rangle$. For that purpose, we split $\zeta$ into a large-scale, classical, background part $\zl$ and a small-scale quantum part $\zs$:
 \be
 \zl \equiv \int_{k < k_*} \frac{d k^3}{(2\pi)^3} \zeta_{\bk}e^{i \bk \cdot \bx}\,, \qquad  \zs \equiv \int_{k > k_*} \frac{d k^3}{(2\pi)^3} \zeta_{\bk}e^{i \bk \cdot \bx}
 \ee
(where $k_*$ is chosen such that $k_3 < k_* \ll k_1 \simeq k_2$), so that $\zeta=\zl+\zs$. Introducing the latter equation into the second- and third-order action \refeq{actionQuad} and \refeq{action3} respectively, we keep the terms of order $\zs^2$ (coming from the second-order action) and $\zs^2 \zl$ (coming from the third-order action). The terms of order $\zs^2$ provide the equations of motion for $\zs$ that determine the free mode functions while the terms of order $\zs^2 \zl$ are treated as perturbations that enable one to compute $\langle \zeta_{\bkone} \zeta_{\bktwo} \rangle_{\bkthree}$ using the Keldysh-Schwinger formalism (other terms in the third-order action like $\zs \zl^2$ do not contribute to this correlation function). 

We now want to stress an important point : the consistency relation \refeq{consistency} relates the bispectrum to the scalar spectral index only for very squeezed triangles, \textit{i.e.} when $k_3 \to 0$ (see the discussions in \cite{Creminelli:2004yq,Chen:2010xk} for instance). In this limit, the quantum to classical transition for the long wavelength mode is pushed away to the past infinity. Therefore, even in cases where large interactions occur while the short-wavelength modes are under the horizon \cite{Chen:2006xjb,Chen:2008wn}, such as in models with features, the long-wavelength mode can be treated as classical and constant. Hence, although the Keldysh-Schwinger formalism involves an integral from early time to the time $\tb$ at which we evaluate the bispectrum, one can neglect, in the limit $k_3 \to 0$, the interactions terms containing time derivatives of the long wavelength mode, as well as terms containing its spatial gradients obviously\footnote{One should be cautious with terms in ${\dot \zs} \zeta_{s,i} \partial^{-2} {\dot \zeta_{l,i }}$ or $(\partial^2 \zs)( \partial^{-2}   {\dot \zeta_{s,i }})(  \partial^{-2}   {\dot \zeta_{l,i }} )$, that do arise in our calculation, where the spatial and time derivatives act on $\zl$ in opposite ways. One can check that these terms are indeed negligible in the squeezed limit (the case of the first one is treated in \cite{Ganc:2010ff}). That would not be the case for instance for terms in ${\dot \zs }^2 \partial^{-2} {\dot \zl} $, absent in our calculation.}. This will enable us to drop a number of terms.

\subsection{The action for short-wavelength modes in the long-wavelength mode background}

Inserting $\zeta=\zl+\zs$ into the second-order action \refeq{actionQuad}, the second-order action for the short-wavelength part $\zs$ is straightforwardly derived:
\be
S_{s^2} = \int dt d^3x~
[a^3  \frac{\epsilon}{c_s^2}\dot\zs^2- a \epsilon
(\partial \zs)^2 ] \,.
\label{2nd-action-zetas}
\ee
Similarly, at zeroth order in the squeezed limit, \textit{i.e.} neglecting any time and space derivative of $\zl$, the terms of order $\zs^2 \zl$ in \refeq{action3} are found to be
\begin{eqnarray}
 \label{action-couplage}
S_{s^2l}&=&\int dt d^3x\left[ \frac{a^3\epsilon}{c_s^4}(\epsilon-3+3c_s^2)\zl \dot{\zs}^2
\right.
\cr
&&
\left.
+\frac{a\epsilon}{c_s^2}(\epsilon-2s+1-c_s^2)\zl (\partial\zs)^2+\frac{a^3\epsilon}{c_s^2}    \Tdot{\left(\frac{\et}{\c^2}\right)} \zl \zs\dot{\zs}
\right.
\cr
&&
\left.
-\left(\frac{\eta}{2 \cs^2} \zl \zs +\frac{1}{\cs^2 H}\zl \dot \zs \right) \frac{\delta S_2}{\delta\zs}  \right] \,.
\end{eqnarray}
The last terms in $\frac{\delta S_2}{\delta\zs}$ is most efficiently treated by using a field redefinition
\be
\zs=\zn+\frac{\eta}{\cs^2} \zl \zn+\ldots
\label{redefinition-explicit}
\ee
where we have omitted the term with a time derivative of the short wavelength mode since this field redefinition will only be evaluated at the time $\tb$ where all the modes have become constant. The second-order action for the redefined field $\zn$, that we call $S_0$, and the cubic interaction action between $\zl$ and $\zn$, that we call $S_{\rm int,(3)}$, then take the form
\bea
\label{2nd-action-zetan}
S_{0} &=& \int dt d^3x~
[a^3  \frac{\epsilon}{c_s^2}\dot\zn^2- a \epsilon
(\partial \zn)^2 ] \,, \nn \\
S_{\rm int,(3)}&=&\int dt d^3x\left[ \frac{a^3\epsilon}{c_s^4}(\epsilon-3+3c_s^2)\zl \dot{\zn}^2+\frac{a\epsilon}{c_s^2}(\epsilon-2s+1-c_s^2)\zl (\partial\zn)^2
\right.
\cr
&&
\left.
+\frac{a^3\epsilon}{c_s^2}    \Tdot{\left(\frac{\et}{\c^2}\right)} \zl \zn\dot{\zn}
  \right]\,.
\label{Sint}
\eea
Note that the vertex in $\dot \zeta^3$ in \refeq{action3}, absent in models with standard kinetic terms, and which generates equilateral type non-Gaussianities, is manifestly irrelevant in the squeezed limit. The non-canonical structure of the action actually only manifests itself in the coefficients of the vertex in \refeq{Sint} being different from the canonical case.

Note that we are eventually interested in the two-point correlation function of $\zs$, and not of $\zn$. From \refeq{redefinition-explicit}, the link between the two is found to be (this is worked out in details in \cite{Ganc:2010ff} for $\cs=1$)
\be
\langle \zeta_{s,\bkone}(\tb)  \zeta_{s,\bktwo}(\tb)  \rangle \simeq \langle \zeta_{n,\bkone}(\tb)  \zeta_{n,\bktwo}(\tb)  \rangle+2 \frac{\eta(\tb)}{\cs^2(\tb)}P_{\zeta}(k_1) \zeta_{l,\bkone+\bktwo}
\label{2-contributions}
\ee 
In the next subsection, we will use the cubic action \refeq{Sint} to compute $\langle \zeta_{n,\bkone}(\tb)  \zeta_{n,\bktwo}(\tb)  \rangle$ in the extreme squeezed limit.

\subsection{Quantizing $\zn$ and applying the Keldysh-Schwinger formalism}
\label{general-for}

We now follow the standard procedure to proceed to the quantification of $\zn$ (see \cite{MFB}, or \cite{Langlois:2010xc}). We first expand $\zn$ in Fourier space
\be
\zn(\bx)=\int \frac{d^3 k}{(2\pi)^3}\zeta_{n,\bk} e^{i \bk \cdot \bx}
\label{Fourier}
\ee
and promote $\zeta_{n,\bk}$ to a quantum operator
\be
\zeta_{n,\bk}(t)=u_k(t){\hat a}_{\bk}+u^*_k(t){\hat a}^{\dagger}_{-\bk}
\label{operator}
\ee
where ${\hat a}_{\bk}$ and ${\hat a}^{\dagger}_{\bk}$ are annihilation and creation operators that satisfy the canonical commutation rules
\be
[{\hat a}_{\bk},{\hat a}^{\dagger}_{\bk^{\prime}} ]
=(2\pi)^3\delta^{(3)}(\bk-\bk^{\prime}).
\label{commutation}
\ee
The so-called mode functions $u_k(t)$ (as well as $u_k^*(t)$) satisfy the classical equations of motion derived from \refeq{2nd-action-zetan}
\be
\frac{\partial}{\partial t} \left(a^3\frac{\eps}{\cs^2}{\dot u_k} \right) +a \eps k^2 u_k=0\,.
\label{eq-mode-function}
\ee
Equivalently, introducing the canonically normalized field in conformal time $\tau=\int dt/a(t)$
\be
v_k \equiv z u_k\,, \qquad z \equiv \frac{a\sqrt{2\eps}}{\cs}\,,
\label{v-u}
\ee
equation \refeq{eq-mode-function} can be recast in the familiar form (a prime denotes a derivative with respect to $\tau$)
\be
v_k^{\prime \prime}+\left( \cs^2 k^2 -\frac{z^{\prime \prime}}{z}\right)v_k=0\,
\label{equation-v}
\ee
where $v_k$ has to satisfy the quantization (Wronskian) condition
\be
v_k^* v_k^{\prime}-v_k v_k^{\prime *}=-i\,
\label{Wronskian}
\ee
and the appropriate vacuum condition (we will treat the case of the standard Bunch-Davies vacuum in the following section).

From \refeq{operator} and \refeq{commutation}, one deduces the power spectrum of $\zn$ (which is the same as the power spectrum of $\zeta$):
\be
 \langle \zeta_{\bk}(t)  \zeta_{\bk^{\prime}}(t) \rangle=(2\pi)^3 \delta^{(3)}(\bk+\bk^{\prime})  |u_{ k}(t) |^2
\ee
so that 
\be
P_{\zeta}(k)=\lim_{aH \gg k c_s}  |u_{ k} |^2 \,.
\ee

We now turn to the two point correlation function of $\zn$ induced by the cubic interactions in \refeq{Sint}
\be
 \langle \zeta_{n,\bkone}(\tb)  \zeta_{n,\bktwo}(\tb) \rangle_{\zeta_{l,\bkthree}}
\ee
 when $\bkone$ and $-\bktwo$ are different (although very close). At tree-level in the Keldysh-Schwinger formalism \cite{Keldysh:1964ud,Schwinger:1960qe}, this is given by \cite{Weinberg:2005vy}
\be
 \langle \zeta_{n,\bkone}(\tb)  \zeta_{n,\bktwo}(\tb) \rangle_{\zeta_{l,\bkthree}}=-i \int_{-\infty}^{\tb} d t \langle 0  | \zeta_{n,\bkone}(\tb)  \zeta_{n,\bktwo}(\tb) H_{I,(3)}(t)  |0 \rangle +{\rm c.c.}
 \label{in-in}
\ee
where $H_{I,(3)}=-L_{\rm int,(3)}$ is the cubic order interaction Hamiltonian and all fields are in the interaction picture (which means that they are free fields). Inserting the expression \refeq{Sint} into \refeq{in-in}, one finds (see appendix \ref{in-in-general-calculation} for the details of the calculation):
\be
 \langle \zeta_{n,\bkone}(\tb)  \zeta_{n,\bktwo}(\tb) \rangle_{\zeta_{l,\bkthree}}=\F \zeta_{l,\bkone+\bktwo}
\ee
with
\bea
\F&=&i u_{k_1}^2({\bar \t}) \int_{-\infty}^{{\bar \t}} d \tau \left[  \frac{2 \e}{\c^4} (\e-3+3 \c^2) a^2 (u_{k_1}^{\prime*})^2 +\frac{2\e}{\c^2}(1-\c^2+\e-2s)a^2 k_1^2(u_{k_1}^*)^2   \right.
 \cr
 && \left. 
+ \frac{2 \e}{\c^2}    \Tdot{\left(\frac{\et}{\c^2}\right)} a^3 u_{k_1}^{\prime*} u_{k_1}^*\right] +{\rm c.c.}\,.
\label{F}
\eea
Using this result together with \refeq{2-contributions}, and correlating with $\zeta_{l,\bkthree}$ as announced, one finds
\be
\lim_{k_3 \to 0}\langle \zeta_{\bkone} \zeta_{\bktwo} \zeta_{\bkthree} \rangle=(2\pi)^3 \delta^{3} (\sum_i \bk_i)P_{\zeta}(k_3)(P_{\zeta}(k_1) \frac{\eta(\tb)}{\c^2(\tb)}+\F)\,.
\label{general-result}
\ee
Equations \refeq{F} and \refeq{general-result} provide an expression for the bispectrum in the (extreme) squeezed limit in terms of the mode functions of $\zeta$ in general Lorentz-invariant single field inflation, the case of a canonical scalar field studied in \cite{Ganc:2010ff} being recovered when $\cs=1$ and $s=0$. We stress that the only approximation that we used in deriving them is to work at zeroth order in the squeezed limit. The various parameters $\eps, \eta \ldots$ entering into \refeq{F}, \refeq{general-result} are only short-hand notations and are therefore neither necessarily small nor slowly varying. The vacuum state is also left arbitrary.

\section{V\lowercase{erifying the consistency relation}}
\label{verification}

In the previous section, we derived an expression \refeq{F}, \refeq{general-result} for the squeezed limit of the bispectrum in general single field inflation. We now use this result to explicitly verify the consistency relation \refeq{consistency} in various cases. We begin by considering an exactly solvable class of models in subsection \ref{exact}. We then move to the main calculations of this paper: the verification of Maldacena's consistency relation at first order in the slow-varying approximation in general single field inflation and at second order in the slow-varying approximation in canonical single field inflation.

\subsection{An exactly solvable class of models : power-law inflation with a constant speed of sound}
\label{exact}

In this subsection, we consider power-law inflation with a constant, but otherwise arbitrary, speed of sound, \textit{i.e.} an inflationary phase with parameters $\epsilon$ ($< 1$ in order to realize inflation) and $\cs$ constant (and hence $\eta=s=0$). In appendix \ref{constant-cs}, we give an example of a model, considered in \cite{Spalinski:2007un,Kinney:2007ag}, that realizes such a scenario with a Dirac-Born-Infeld type of Lagrangian. Note however that the proof of the consistency relation below does not depend on the details of such a realization.\\

From $\eps={\rm constant}$, one deduces that $a \propto (t-t_0)^{1/\epsilon}$ and hence
\be
a(\t) =\left( -\tau \right)^{-\frac{1}{1-\epsilon}}\,
\label{a-exact}
\ee
up to a choice of normalization and origin of time. Equation \ref{equation-v} then takes the form
\be
v_k^{\prime \prime}+\left( \cs^2 k^2 -\frac{\nu^2-\frac{1}{4}}{\tau^2}   \right)v_k=0\,
\label{equation-v-Bessel}
\ee
with
\be
\nu=\frac{3-\epsilon}{2(1-\epsilon)}\,.
\label{nu}
\ee
The exact solutions of equation \refeq{equation-v-Bessel}, \refeq{nu} are known, and those with positive frequency modes --- \textit{i.e.} we choose the Bunch-Davies vacuum --- that obey the normalization condition \refeq{Wronskian} read
\be
v_k(\tau)=i \frac{\sqrt{\pi}}{2} \sqrt{-\tau}H_{\nu}^{(1)}(-k \cs \tau)\,
\ee
or equivalently
\be
u_k(\tau)=i \frac{\cs}{2} \sqrt{\frac{\pi}{2 \epsilon}} (-\tau)^{\nu}H_{\nu}^{(1)}(-k \cs \tau)\,
\label{u-exact}
\ee
where $H_{\nu}^{(1)}$ is the Hankel function of the first kind and of order $\nu$  (useful properties of the Hankel functions are collected in the appendix \ref{Hankel}). From the late time behaviour of the Hankel function \refeq{Hankel-0}, one finds
\be
\lim_{\tau \to 0} u_k(\tau)=-\frac{\cs}{2\sqrt{2 \epsilon \pi}}  \Gamma(\nu)\left( \frac{2}{k \cs} \right)^{\nu}
\ee
and hence the scalar spectral index
\bea
n_s-1\equiv  \left. \frac{ d \, {\rm ln }\left( k^3   |u_{ k}(\tau) |^2  \right)}{ d\, {\rm ln} k} \right|_{\tau \to 0} =3-2 \nu =-\frac{2 \epsilon}{1-\epsilon}\,.
\label{ns-exact}
\eea
Now using the relation \refeq{deriv}, one calculates that
\be
u_k^{\prime}(\tau)=-\frac{i \cs^2}{2}\sqrt{\frac{\pi}{2 \epsilon}} k (-\tau)^{\nu}H_{\nu-1}^{(1)}(-k \cs \tau)\,.
\label{u'-exact}
\ee
Plugging \refeq{a-exact}, \refeq{u-exact} and \refeq{u'-exact} into our general result \refeq{F}, one then finds
\be
\frac{\F}{\Pz(k_1)}= -\frac{i\pi}{4 \cs^2} \left[     \left(\epsilon-3+3 \cs^2\right)\int_{-k_1 \cs {\bar \t}}^{\infty} d x x   \left( H_{\nu-1}^{(2)}(x) \right)^2  
 +\left(1-\cs^2+\epsilon \right)   \int_{-k_1 \cs {\bar \t}}^{\infty} d x x   \left( H_{\nu}^{(2)}(x) \right)^2  \right]+{\rm c.c.}
\ee
Using formulas given in appendix \ref{Hankel}, one gets
\be
i  \int_{0}^{\infty} dx\, x   \left( H_{\nu}^{(2)}(x) \right)^2 +{\rm c.c.}=-\frac{4}{\pi} \nu\,,
\ee
so that 
\be
\frac{\F}{\Pz(k_1)}=\frac{\epsilon}{\cs^2}(2\nu-1)+\invc (3-2\nu)\,.
\ee
With $\nu$ given in \refeq{nu}, one sees that the terms with negative powers of $\cs$ disappear and one is left with
\be
\frac{\F}{\Pz(k_1)}=\frac{2 \epsilon}{1-\epsilon}\,.
\ee
It is then clear from equations \refeq{general-result} and \refeq{ns-exact} that we have verified the consistency relation \refeq{consistency} in this exactly solvable class of models with an arbitrary speed of sound, thereby generalizing the results obtained in \cite{Ganc:2010ff} for canonical power law inflation.

\subsection{Slowly varying general single field inflation}
\label{slow}

In the remainder of this paper, we consider a slowly-varying inflationary phase (though with an arbitrary speed of sound unless otherwise specified) with perturbations in the standard Bunch-Davies vacuum. The consistency relation \refeq{consistency} has already been checked in this general class of models at leading order in the slow-varying approximation in references \cite{Chen:2006nt} and \cite{Cheung:2007sv}. Although our calculation obviously shares common features with the ones in these papers, we stress that it is largely different. Indeed, Chen \textit{et al.} \cite{Chen:2006nt} calculated the full bispectrum through a complete quantum calculation, and later took the squeezed limit, whereas we consider this limit directly, resulting in a considerable simplification. This latter approach is also followed by Cheung \textit{et al.} but in the framework of the effective field theory of inflation \cite{Cheung:2007st}, and in practice with a completely different method. Moreover, we will show that our approach enables one to verify quite readily Maldacena's consistency relation \textit{at second order} in the slow-varying approximation in models with canonical kinetic terms, a new result to the best of our knowledge.

By a slowly-varying inflationary phase, we mean that the parameters $\eps,\eta$ and $s$, that were arbitrary for the moment, are considered both as much smaller than unity, which we note as $\CO(\eps)$, and slowly varying, \textit{i.e.} $\frac{1}{H}\left({\dot \eps},{\dot \eta},{\dot s} \right)= \CO(\eps^2)$. In the following, we refer to an expression as being of the $n$-th order (in the slow-varying approximation) when it is accurate up to $\CO(\eps^n)$ terms. As for the speed of sound, we remind the reader that although it equals one in canonical single field inflation, there are known examples, like DBI inflation \cite{DBI1,DBI2}, where it can be much less than unity. In the following, we hence leave it arbitrary.

For our purpose, we will need the solution to the mode equation \ref{equation-v} (and that verifies the normalization condition \refeq{Wronskian}) up to first order in the slow-varying approximation. This is given by\footnote{The expression given here differs from the one in \cite{Chen:2006nt} by a phase factor which can be considered as constant at this order and which therefore is irrelevant.} (we refer the reader to \cite{Chen:2006nt} for an explicit derivation)
\bea
u_k(y) = i \frac{\sqrt{\pi}}{2\sqrt{2}} ~\frac{H}{\sqrt{\epsilon c_s}}~
\frac{1}{k^{3/2}} (1+\frac{\epsilon}{2} +\frac{s}{2} )
~ y^{3/2}
H_\nu^{(1)} \left( (1+\epsilon+s)y \right) \left( 1+\CO(\epsilon^2)\right) 
\label{ukepsilon}
\eea
where we chose the Bunch-Davies vacuum and we defined
\bea
y \equiv \frac{k \cs}{a H}\,
\eea
and
\bea
\nu \equiv \frac{3}{2} + \epsilon + \frac{\eta}{2} +\frac{s}{2} ~.
\eea
Note that at zeroth order in the slow-varying approximation, in which $a=-\frac{1}{H \tau}(1+\CO(\eps))$ and $\nu =\frac{3}{2}+\CO(\eps)$, expression \refeq{ukepsilon} simplifies to the well known result

\begin{eqnarray}
u_k(\tau)& =& \frac{H_K}{\sqrt{4\epsilon_K
c_{sK} k^3}}(1+i k c_{sK}\tau)e^{-i k c_{sK}\tau}  \, \qquad 0^{th }\,{\rm order }
\label{uk}  \\
u_k^{\prime}(\t)&=&\frac{H_K}{\sqrt{4 \epsilon_K c_{sK} k^3}} k^2 c_{sK}^2 \t e^{-i k c_{sK} \t}   \, \qquad \,\,\,\,\,\, \,\,\,\,\,\,\,\, 0^{th }\,{\rm order }\,
\label{uk'}
\end{eqnarray}
where all parameters are considered as constant at this order and we have chosen to evaluate them at the arbitrary pivot point $\tau_K$ (though near $\tau_k$ for consistency) for $K=\a k$. Here and in the following, a subscript $K$ indicates that the corresponding quantity is evaluated at sound horizon crossing $K c_{sK}=a_K H_K$.

From the full result \refeq{ukepsilon}, one deduces the asymptotic value of the mode function up to $\CO(\eps)$ order \cite{Chen:2006nt}:
\bea
u_k(0) = \frac{ H_k}{2\sqrt{c_{s k}\epsilon_k}} \frac{1}{k^{3/2}}
\left(1-(\bc+1)\epsilon - \frac{\bc}{2}\eta -\left(\frac{\bc}{2} +1\right)s
\right)  
\label{uk0corr}
\eea
where 
\be
\bc \equiv \gamma-2+\ln 2 \approx -0.73 \,,
\label{C}
\ee
and $\gamma = 0.577\cdots$ is the Euler constant. With our convention, the mode functions become real at late times so that 
\bea
u_k^2(0)=| u_k^2(0) |=  P_{\zeta}(k)
\label{u-reel}
\eea
and hence, with $  \frac{ \left. d\,   f(t) \right|_{t_k}}{ d\, {\rm ln} k}=  H_k^{-1}(1-\epsilon_k)^{-1} {\dot f}(t_k)  $, one obtains the scalar spectral index at second order in the slow varying approximation:
\bea
n_s(k)-1&=&-2 \eps_{k}-\eta_{k}-s_{k} - 2 \eps_k^2-\eta_k \eps_k -s_k \eps_k\nn \\
&  -& 2(\bc+1)\eta_k \eps_k -\bc \frac{\dot \eta_k}{H_k \eta_k} \eta_k
 -(\bc+2) \frac{\dot s_k}{H_k s_k} s_k
+\CO(\eps^3)\,.
\label{ns}
\eea

Finally, inserting \refeq{u-reel} into the expression \refeq{F}, the latter becomes
\bea
\F &=&\F_1+\F_2+ \F_3 \qquad {\rm with}
\label{F-sum} \\
\F_1&=&4 P_{\zeta}(k_1)\, {\rm Re} \left[  -i \int_{-\infty}^{{\bar \t}} d \tau  g_1 (\t)  a^2 (u_{k_1}^{\prime*})^2   \right]
\label{F1} \\
\F_2&=&4 P_{\zeta}(k_1) \,{\rm Re} \left[  -i \int_{-\infty}^{{\bar \t}} d \tau  g_2 (\t)  a^2 k_1^2 (u_{k_1}^*)^2   \right] 
\label{F2} \\
\F_3&=&4 P_{\zeta}(k_1) \,{\rm Re} \left[  -i \int_{-\infty}^{{\bar \t}} d \tau  g_3 (\t)  a^3  u_{k_1}^{\prime*} u_{k_1}^*    \right] 
\label{F3}
\eea
where
\bea
g_1(\t)&=& \frac{\e}{\c^4} (3-3 \c^2-\e) 
\label{g1}\\
g_2(\t)&=&-\frac{\e}{\c^2}(1-\c^2+\e-2s)
\label{g2}\\
g_3(\tau)&=&-\frac{\epsilon}{\cs^2}   \Tdot{\left(\frac{\eta}{\cs^2}\right)} \,.
\label{g3} 
\eea
In the following, when evaluating \refeq{F1}, \refeq{F2} and \refeq{F3}, we simply note $k=k_1$ for brevity. We note also that $\F_3$ is $\CO(\eps)$ smaller than $\F_1$ and $\F_2$ and hence can be neglected at leading order in the slow-varying approximation.

\subsubsection{A warm-up: canonical inflation at leading order in the slow-varying approximation}
\label{warm-up}

As a warm-up, we consider the case of a canonical inflationary Lagrangian, for which $\cs=1$ and $s=0$, at leading order in the slow-varying approximation. Using \refeq{uk} and \refeq{uk'}, one finds, at zeroth order in the slow-varying approximation
\bea
\frac{\F_1}{\Pz(k)}&=&-\eps_K \, \k {\rm Re} \left[-i \int_{-\infty}^{\bar \t} d \t e^{2i k \t} \right] \nn \\
\frac{\F_2}{\Pz(k)}&=&  -\frac{\eps_K}{k}   {\rm Re} \left[-i \int_{-\infty}^{\bar \t} \frac{d \t}{\t^2}(1-i \k \t)^2 e^{2i \k \t} \right]\,.
\eea
As usual in such calculations, the integrals that are required are regulated in the infinite past by using the appropriate contour in the complex plane $\t \to -(\infty-i\eps)$ and their integration bound is extrapolated to $\tau=0$ as most of their contributions comes from the period around horizon crossing \cite{Weinberg:2005vy} (we will treat a case where this is not true in subsection \ref{canonical-2}). Using the integrals given in appendix \ref{integrals} (where we collect the integrals that are needed here and in the following), one finds
\bea
\F_1+\F_2&=&2 \eps_K \Pz(k)\,.
\label{F-simple}
\eea
Remember also that the time $\bar \tau$ is chosen such that the mode functions have then reached their asymptotic value, typically a few efolds after sound horizon crossing. Hence, at leading order in the slow-varying approximation, the parameter $\eta(\bar \tau)$ appearing in the general result\refeq{general-result} can be considered as being equal to $\eta_K$ (this will not remain true at next to leading order as we will see). From \refeq{F-simple} and \refeq{general-result}, we thus find
\bea
\lim_{k_3 \to 0} \langle \zeta_{\bkone} \zeta_{\bktwo}\zeta_{\bkthree} \rangle &=&- (2 \pi)^3  \delta^{(3)} (\sum_i \bk_i)P_{\zeta}(k_1)P_{\zeta}(k_3)  \left( -2 \eps_K -\eta_K +\CO(\eps^2) \right) \nn \\
&=&- (2 \pi)^3  \delta^{(3)} (\sum_i \bk_i)P_{\zeta}(k_1)P_{\zeta}(k_3)  \left( -2 \eps_{k_1} +\eta_{k_1} -\CO(\eps^2) \right) \,.
\eea
It is then clear from the expression of the scalar spectral index \refeq{ns} that the consistency relation \refeq{consistency} is indeed verified.

\subsubsection{The case of an arbitrary speed of sound at leading order in the slow-varying approximation}
\label{general-first-order}

We now move on to the case where the speed of sound is arbitrary, for which we will see that verifying the consistency relation at the first non-trivial order requires much more work. Indeed, given the forms of the coefficients $g_1$ and $g_2$ in \refeq{g1} and \refeq{g2} and of the solutions \refeq{uk} and \refeq{uk'}, one expects from the previous analysis the right-hand side of the relation \refeq{general-result} to have contributions of order $\CO(\frac{1}{\cs^2})$ and $\CO(\frac{\eps}{\cs^2})$ (not including the factor $\Pz(k_1) \Pz(k_3$)), whereas the scalar spectral index \refeq{ns} starts at order $\CO(\eps)$. To check the consistency relation at this order, one therefore has to verify that the terms evolving negative powers of $c_s$ disappear as well as to take into account all the $\CO(\eps)$ corrections to this naive reasoning.

Working for the moment at leading order in the slow-varying approximation requires the same type of integrals as in the previous subsection, the only difference being the coefficients:
\bea
\frac{\F_{\rm naive}}{\Pz(k)}&=&\left[ 3(1-\c^2)-\e\right]\frac{\k}{\c}{\rm Re} \left[-i \int_{-\infty}^0 d \t e^{2i \k \c \t} \right] \nn \\
& -&\frac{1}{\c^2}(1-\c^2+\e-2s)\frac{1}{\c \k}   {\rm Re} \left[-i \int_{-\infty}^0 \frac{d \t}{\t^2}(1-i \k \c \t)^2 e^{2i \k \c \t} \right] \nn \\
&=&\frac{2 \e-3s}{\c^2}
\label{naive}
\eea
where, from now on, we omit the subscript $K$ when the context is clear. Hence we have verified that the leading order terms in $\CO(\frac{1}{\cs^2})$ disappear. However, the calculation is not consistent at this stage because by treating all the slow-varying parameters in the integrands as constant, we neglected $\CO(\epsilon)$ corrections, which, multiplied by the $1/\cs^2-1$ factor in $g_1$ and $g_2$, compete with the result \refeq{naive}. 

To proceed further, we thus need to expand the various functions in \refeq{F1}, \refeq{F2} around the time of sound horizon crossing for our pivot scale. However, the spectral index being at least of order $\CO(\eps)$, the precise value of this pivot scale should be irrelevant if one wants to to check the consistency relation up to order $\CO(\eps)$. The fact that $\a$ will disappear from our final result therefore provides a useful check of our calculation. \\

There are three types of $\CO(\eps)$ corrections to the integrands in \refeq{F1} and \refeq{F2} that need to be taken into account. They were already given in \cite{Chen:2006nt} and we list them below while referring the reader to \cite{Chen:2006nt} for explicit derivations.

\begin{itemize}

\item Corrections to the scale factor:
\bea
a(\t) = -\frac{1}{H_K \tau} - \frac{\epsilon}{H_K \tau} 
+ \frac{\epsilon}{H_K \tau} 
\ln (\tau/\tau_K) + \CO(\epsilon^2) \,.
\label{corr-a}
\eea

\item Corrections to the coupling constants:
\bea
g(\t) &=& g(t_K) + \frac{d g}{d t} (t-t_K) +
\CO (\epsilon^2 g) \nonumber \\
&=& g(\tau_K) - \frac{d g}{d t} \frac{1}{H_K} \ln
\frac{\tau}{\tau_K} + \CO(\epsilon^2 g) \,
\label{corr-g}
\eea
where $g$ collectively stands for $g_1$ and $g_2$.

\item Correction to the mode functions:

Denoting as $\Delta u_{\k}(\tau)$ and $ \Delta u_{\k}^{\prime }(\tau)$ the $\CO(\eps)$ corrections to respectively \refeq{uk} and \refeq{uk'}, we obtain from \refeq{ukepsilon}
\bea
\Delta u_k^*(\tau) &=& -\frac{i}{2} 
\frac{H_K}{\sqrt{c_{s K}\epsilon_K}} \frac{1}{\k^{3/2}} e^{-ix} 
\nonumber \\
&\times& \left[(\epsilon + s) (x-i) + i s x^2 +\left(-(\epsilon +\frac{\eta}{2} +\frac{s}{2})(x-i)  - i x^2 s\right) \ln\frac{\tau}{\tau_K}
\nn
\right.
\cr
&&
\left.
+   (\epsilon +\frac{\eta}{2}
+\frac{s}{2})e^{ix} h^*(x) \right] 
\label{Du*}
\eea
and
\bea
 \Delta u_{\k}^{\prime *}(\tau) &=& 
\frac{i}{2} \frac{H_K}{\sqrt{c_{s K}\epsilon_K}} \frac{1}{\k^{3/2}}
\k  c_{s K} e^{-ix} \nn \\
&\times&
 \left[ -(\epsilon +\frac{\eta}{2} +\frac{s}{2})(1-\frac{i}{x}) - i\epsilon x + s x^2 
+ ( i\epsilon +\frac{i}{2} \eta - \frac{3}{2}i s - s x) x
\ln\frac{\tau}{\tau_K} 
\nonumber 
\nn
\right.
\cr
&&
\left.
+ (\epsilon +\frac{\eta}{2} +\frac{s}{2})e^{ix} 
\frac{d h^*(x)}{dx}
\right] 
\label{dDu*dtau}
\eea
where $x\equiv -\k c_{s K} \tau$ and we defined
\be
h(x) \equiv  \sqrt{ \frac{\pi}{2}} x^{3/2} \left[ \frac{d H^{(1)}_\nu(x)}{d\nu} \right]_{\nu=\frac{3}{2}}
\ee
(explicit expressions of $h^*(x)$ and its derivative in terms of special functions are given in the appendix \ref{correction-details}).

\end{itemize}

We then consider all three types of corrections to the two integrations \refeq{F1} and \refeq{F2}. We give the details of these long calculations in appendix \ref{correction-details} and simply give the final result here:
\bea
\frac{\eta(\bar \tau)}{\cs^2(\bar \tau)}+\frac{\F}{\Pz(k)}&=& \left( \frac{\eta_K}{c_{sK}^2}+\frac{\F_{\rm naive}}{\Pz(k)}+\frac{\Delta \F}{\Pz(k)} \right) \left( 1+\CO(\eps)\right) \nn \\
&=& \left(  \frac{\eta_K}{c_{sK}^2}+\frac{2 \eps_K-3s_K}{c_{sK}^2}  -\left( \frac{1}{c_{sK}^2}-1\right) (2 \eps_K+\eta_K-3s_K)+4s_K \right) \left(1+\CO(\eps) \right) \nn \\
&=&2\eps_K+\eta_K+s_K +\CO(\eps^2) \nn \\
&=&2\eps_k+\eta_k+s_k +\CO(\eps^2)
\eea
where we added the term coming from the field redefinition in \refeq{general-result}, $\F_{\rm naive}/\Pz(k)$ was already given in \refeq{naive} and $\Delta \F/\Pz(k)$ denotes the contribution induced by the corrections that we listed above. From the expression of the scalar spectral index \refeq{ns}, it is thus clear that we have verified the consistency relation
\be
\lim_{k_3 \to 0} \langle \zeta_{\bkone} \zeta_{\bktwo}\zeta_{\bkthree} \rangle = -(2 \pi)^3  \delta^{(3)} (\sum_i \bk_i)(n_s(k_1)-1)P_{\zeta}(k_1)P_{\zeta}(k_3) 
\ee
up to $\CO(\eps)$ order.

\subsubsection{Canonical inflation at next to leading order in the slow-varying approximation}
\label{canonical-2}

We now show that one can verify Maldacena's consistency relation at second order in the slow-varying approximation in standard single field inflation with surprisingly few efforts given our previous work. First, one needs to compute the $\CO(\eps)$ corrections to the result \refeq{F-simple} for $\F_1+\F_2$, which can be straightforwardly derived from the calculations of the last subsection. Second, one has to evaluate $\F_3$ \refeq{F3}, but being suppressed by $\CO(\eps)$ terms with respect to $\F_1$ and $\F_2$, at leading order only. Finally, one has to take into account the fact that the term coming from the field redefinition in \refeq{general-result} is evaluated at the late time $\bar \tau$, and not at horizon crossing. We now treat these three calculations successively.

\begin{itemize}

\item Corrections to $\F_1+\F_2$ \refeq{F-simple}:

For our calculation in subsection \ref{warm-up}, evaluating $\F_1+\F_2$ at zeroth-order in the slow-varying approximation was sufficient to obtain the required result. At next to leading order however, one must take into account all the $\CO(\eps)$ corrections to the naive behaviour of the scale factor, the mode functions and the coupling constants. This is exactly what we have done in the last subsection in the general single field case!, with the only difference that $g_1(\tau)= \frac{\e}{\c^4} (3-3 \c^2-\e)$ and $g_2(\t)=-\frac{\e}{\c^2}(1-\c^2+\e-2s)$ now take their ``canonical'' value when $\cs=1$, \textit{i.e.} $-\eps^2$. Hence, no new calculations are required and the results can simply be deduced from the integrals in appendix \ref{correction-details} by changing their multiplicative coefficients: the corrections to $\F_1$ (respectively to $\F_2$) coming from the scale factor and from the mode functions are obtained from the results \refeq{DaF1} and \refeq{DuF1} (respectively \refeq{DaF2} and \refeq{DuF2}) by making the replacement $1/\cs^2-1 \to -\eps/3$ (respectively $1/\cs^2-1 \to \eps$). As for the correction to $\F_1$ (respectively to $\F_2$) coming from the time variation of the coupling constant, it can be deduced from \refeq{DgF1} (respectively from \refeq{DgF2}) by making the replacement $-3 \left(\frac{1}{\c^2}-1\right)(\et-4s)  +6 s  \to 2 \eps \eta$ (respectively $  \left(\frac{1}{\c^2}-1\right)(\et-2s)  -2 s \to 2 \eps \eta$). Summing all these contributions, one finds

\be
\frac{\Delta \F_1}{\Pz(k)}+\frac{\Delta \F_2}{\Pz(k)}= 2\eps_K^2 +\eps_K \eta_K \left( -1+2\left(\gamma-\l \right)   \right) \,.
\ee

\item Calculation of $\F_3$:

Inserting the expressions at leading order for the scale factor $a=-\frac{1+\CO(\eps)}{H_K \tau}$, the coupling constant \refeq{g3} and the mode functions \refeq{uk}, \refeq{uk'} into \refeq{F3} for $\cs=1$, one finds

\bea
\frac{\F_3}{\Pz(k)}=  \left( \frac{{\dot \eta}_K}{H_K \eta_K} \right) \eta_K \, \frac{1}{k}  \, {\rm Re} \left[-i \int_{-\infty}^{\bar \t} \frac{d \t}{\t^2}(1-i \k \t) e^{2i \k \t} \right]\,.
\eea
Here, one can not simply extrapolate the integration bound to $\tau=0$ because the integral that is required does not converge. We instead have to keep the $\bar \tau$ dependence of the integral:
\bea
\frac{\F_3}{\Pz(k)}=  \left( \frac{{\dot \eta}_K}{H_K \eta_K} \right) \eta_K \, \left( - \frac{{\rm sin}(2 k {\bar \tau})}{k {\bar \tau}}+ {\rm Ci}(-2k {\bar \tau})  \right)\,
\eea
where 
\bea
{\rm Ci}(x)&=&-\int_x^{\infty} dt\, \frac{{\rm cos}(t)}{t} \, \qquad (x > 0) \nn \\
&=&\gamma+{\rm ln}x +\CO(x^2)\,.
\eea
Hence, one obtains
\be
\frac{\F_3}{\Pz(k)}=  \left( \frac{{\dot \eta}_K}{H_K \eta_K} \right) \eta_K \,   \bc + \left( \frac{{\dot \eta}_K}{H_K \eta_K} \right) \eta_K\, {\rm ln}(-k {\bar \tau})
\ee
up to terms that are negligible for $\bar \tau \to 0$ and where the numerical factor $\bc$ was already defined in \refeq{C}.

\item Corrections to the naive result from the field redefinition:

Using \refeq{corr-g}, one relates $\eta(\bar \tau)$ to $\eta_K$ at next to leading order in the slow-varying approximation:
\be
\eta(\bar \tau) = \eta_K \left(1-  \frac{{\dot \eta}_K}{H_K \eta_K}  {\rm ln}(-K {\bar \tau}) + \CO(\epsilon^2) \right)  \,.     
\ee

\end{itemize}

Adding all the above contributions, one finds that the terms in ${\rm ln}(-k {\bar \tau})$ disappear --- as expected as the bispectrum does not depend on the time at which it is evaluated after all the modes have become constant --- and one finds:
\bea
\eta(\bar \tau)+\frac{\F}{\Pz(k)}&=& 2 \eps_K+\eta_K +2 \eps_K^2+\eps_K \eta_K\left(-1+2\left(\gamma-\l \right) \right)\nn \\
&+& \left( \frac{{\dot \eta}_K}{H_K \eta_K} \right) \eta_K \,   (\bc-{\rm ln}\, \a)+\CO(\eps^3)\,.
\label{sum} 
\eea
Now using that
\bea
\eta_K &=& \eta_{K^{\prime}} \left(1+  \frac{{\dot \eta}_{K^{\prime}}}{H_{K^{\prime}} \eta_{K^{\prime}}}  {\rm ln} \frac{\a}{\alpha_{K^{\prime}}} + \CO(\epsilon^2) \right)    \\
\eps_K &=& \eps_{K^{\prime}} \left(1+  \eta_{K^{\prime}}  {\rm ln} \frac{\a}{\alpha_{K^{\prime}}} + \CO(\epsilon^2) \right)\,,
\label{eps-K-K'}
\eea
it is manifest that the right hand side of \refeq{sum} is independent of the pivot scale $K$ at this order, which provides a useful check of our calculation which would not have been possible with the two ``natural'' choices $K=k$ and $K=2k$. In the former case, the slow-varying parameters appear directly evaluated at the scale of interest at the cost of having a more intricate calculation of the integrals while in the latter case, the time integrals are more straightforwardly performed and taylor expansions such as \refeq{eps-K-K'} are needed in the end. Now expressing the right hand side of \refeq{sum} in terms of parameters evaluated at the scale $k$, one finds
\be
\eta(\bar \tau)+\frac{\F}{\Pz(k)}= 2 \eps_k+\eta_k+2 \eps_k^2+\eps_K \eta_K\left(3+2 \bc \right)   + \left( \frac{{\dot \eta}_k}{H_k \eta_k} \right) \eta_k \,  \bc +\CO(\epsilon^3)   \,.
\label{sum2} 
\ee
Hence it is clear from this expression, the general result \refeq{general-result} and the scalar spectral index \refeq{ns} that we have successfully verified the consistency relation \refeq{consistency} at second order in the slow-varying approximation.

\section{C\lowercase{onclusions}}
\label{conclusion}

In this paper, we have generalized a recent result from Ganc and Komatsu \cite{Ganc:2010ff} by giving an explicit formula for the  squeezed limit of the primordial bispectrum in any Lorentz-invariant model of single field inflation \refeq{F}, \refeq{general-result}. We stress that besides working in the extreme squeezed limit $k_3 \to 0$, no approximation was used, in particular of slowly-varying type, in the derivation of this formula which requires an integral over time involving the free mode functions of $\zeta$. We then used this formula to verify the consistency relation \refeq{consistency} in various specific cases, exemplifying that no approximation was made by considering an exactly solvable class of models, namely power-law inflation with a constant speed of sound.

Specifying then to the slow-varying regime, we were able to verify the consistency relation at the first non-trivial order in general single field inflation. Although this result has already been obtained in references \cite{Chen:2006nt,Cheung:2007sv}, we believe our derivation to be useful. One of the advantage of the approach presented here, shared by the proof in \cite{Cheung:2007sv}, is that the interaction in $\dot \zeta^3$ in the cubic action \refeq{action3}, dominant for generic configurations of the momenta, is irrelevant in the squeezed limit to every order in the slow-varying approximation, and hence can be neglected from the very beginning. We have also checked explicitly that the pivot scale necessary to verify the consistency relation at order $\CO(\eps)$ can be chosen arbitrarily. 

We have finally verified the consistency relation at second order in the slow-varying approximation in canonical single field inflation, again considering an arbitrary pivot scale, which we think to provide a useful consistency check in such calculations. One should also note a subtlety that arises when going beyond leading order in the slow-varying approximation: the verification of the $\bar \tau$ independence of the final result for the bispectrum \refeq{F}, \refeq{general-result}. We have indeed shown that one must take into account the fact that the terms coming from the field redefinition is evaluated at late time, and note at precisely sound horizon crossing, to obtain the correct, meaningful, result.

Let us finally note that in multiple field models, the squeezed limit of the primordial bispectrum is not related in general to the deviation of the curvature perturbation from scale invariance and hence is not necessarily small. If a large bispectrum is detected in the squeezed limit, multiple field models --- together with the large scale nonlinear evolution that they offer \cite{Wands:2010af,Byrnes:2010em,Lehners:2010fy,Bernardeau:2010jp} --- will have to be considered seriously.

\medskip

\begin{acknowledgments}

I would like to thank Eiichiro Komatsu for his encouragement and comments on a draft version of this paper, Xingang Chen for bringing my attention to reference \cite{Cheung:2007sv} as well as Alejandro Boh\'e and David Langlois for useful discussions.

\end{acknowledgments}

\appendix

\section{Details on the calculations in subsection \ref{general-for}}
\label{in-in-general-calculation}

Inserting the Fourier decomposition \refeq{Fourier} into the cubic action \refeq{Sint}, the expression \refeq{in-in} becomes
\bea
 \langle \zeta_{n,\bkone}(\tb)  \zeta_{n,\bktwo}(\tb) \rangle_{\zeta_{l,\bkthree}}&=&i \int_{-\infty}^{\tb} d t \int \frac{d^3 q_1 d^3 q_2}{(2\pi)^6} \zeta_{l,-\bqone -\bqtwo} (t)\times \nn \\
 && \left[     \frac{a^3\epsilon}{c_s^4}(\epsilon-3+3c_s^2) \, \langle 0  | \zeta_{\bkone}(\tb)  \zeta_{\bktwo}(\tb)  \dot{\zeta}_{\bqone}(t)  \dot{\zeta}_{\bqtwo}(t)  |0 \rangle 
  \right.
\cr
&&
\left.
-\frac{a\epsilon}{c_s^2}(\epsilon-2s+1-c_s^2) (\bqone \cdot \bqtwo) \,   \langle 0  | \zeta_{\bkone}(\tb)  \zeta_{\bktwo}(\tb)  \zeta_{\bqone}(t) \zeta_{\bqtwo} (t)   |0 \rangle 
 \right.
\cr
&&
\left.
+\frac{a^3\epsilon}{c_s^2}    \Tdot{\left(\frac{\et}{\c^2}\right)}  \, \langle 0  | \zeta_{\bkone}(\tb)  \zeta_{\bktwo}(\tb)    \dot{\zeta}_{\bqone}(t)  \zeta_{\bqtwo}(t)  |0 \rangle      \right]+{\rm c.c.}
 \label{in-in-explicit}
\eea
where we omit the subscript $n$ on $\zn$ from now on if the context is clear. Using the decomposition into creation and annihilation operators \refeq{operator} together with the commutation rules \refeq{commutation}, one finds
\bea
i   \langle 0  | \zeta_{\bkone}(\tb)  \zeta_{\bktwo}(\tb)  \zeta_{\bqone}(t) \zeta_{\bqtwo} (t)   |0 \rangle&=&i(2\pi)^6 \left( \delta^{(3)} (\bkone+\bqone)  \delta^{(3)} (\bktwo+\bqtwo) +\delta^{(3)} (\bkone+\bqtwo)  \delta^{(3)} (\bktwo+\bqone)   \right)  \nn \\
&\times& \left[ u_{k_1}(\tb) u_{k_2}(\tb) u^*_{q_1}(t) u^*_{q_2}(t)  \right]\,,
\eea
and similarly when one of the $\zeta_{\bqone}(t)$ is replaced by ${\dot \zeta}_{\bqone}(t)$, and where we have dropped a term proportional to $\delta^{(3)} (\bkone+\bktwo) $ because $\bkone \neq -\bktwo$ in our calculation. Assembling everything and working at leading order in the squeezed limit, \textit{i.e.} considering that $\bkone \cdot \bktwo \simeq -k_1^2 \simeq -k_2^2$, one gets
\bea
 &&\langle  \zeta_{n,\bkone}(\tb)  \zeta_{n,\bktwo}(\tb) \rangle_{\zeta_{l,\bkthree}}=i  u_{k_1}^2(\tb) \int_{-\infty}^{\tb} d t \, \zeta_{l,\bkone +\bktwo} (t) \, \times \nn \\
 && \left[     \frac{a^3\epsilon}{c_s^4}(\epsilon-3+3c_s^2) \, 2(u_{k_1}^{\prime*}(t))^2 +\frac{a\epsilon}{c_s^2}(\epsilon-2s+1-c_s^2) 2 k_1^2 \,    (u_{k_1}^{*}(t))^2
  \right.
\cr
&&
\left.
+\frac{a^3\epsilon}{c_s^2}    \Tdot{\left(\frac{\et}{\c^2}\right)}  \, 2 u_{k_1}^{\prime*}(t)  u_{k_1}^{*}(t)   \right]+{\rm c.c.}
 \label{in-in-explicit-2}
\eea
In this kind of integrals that are familiar in the Keldysh-Schwinger formalism, most of the contribution usually come from the period around sound horizon crossing $k \cs \approx aH$ because the rapid oscillations of the mode functions prior to that epoch usually average out. In this case, it is obvious that one can put the term $\zeta_{l,\bkone+\bktwo}$ out of the integral when $k_3 \ll k_1 \simeq k_2$ because $\zl$ has then reached its final constant value since a long time. What are less straightforward are the situations where the time integral in \refeq{in-in-explicit-2} has a non-zero contribution from the epoch where the short wavelength modes are under the sound horizon. This typically arises when there are features in the background evolution so that the factor $\frac{\epsilon}{c_s^2}    \Tdot{\left(\frac{\et}{\c^2}\right)}$ changes suddenly (in less than a Hubble time). In that case, $\zeta_{l,\bkone+\bktwo}$ may well be still in its quantum regime around the time of the feature and a full quantum calculation is required for finite $k_3$ (as well as it is not legitimate to drop terms in ${\dot \zeta}_{l}$ in the interaction action \refeq{action-couplage}). However, because we are interested in this calculation in the limit where the triangle formed by the three wavevectors is infinitely squeezed, \textit{i.e.} $k_3 \to 0$, one can legitimately consider $\zl$ as constant in the integral. One then arrives at the expressions \refeq{F}, \refeq{general-result}.

\section{Power law Dirac-Born-Infeld inflation}
\label{constant-cs}

We give here an example, considered in \cite{Spalinski:2007un,Kinney:2007ag}, of a scalar field Lagrangian as well as a background evolution for which $\eps$ \refeq{eps} and $\cs$ \refeq{cs} are constant. For that purpose, we consider a class of Lagrangian inspired by string theory and known as of Dirac-Born-Infeld type:
\be
P=-\frac{1}{f(\phi)} \left(\sqrt{1-2 f(\phi) X}-1 \right)-V(\phi)\,.
\label{P-DBI}
\ee
When derived from string theory, $\phi$ in \refeq{P-DBI} represents the location of a D3-brane extended along the four usual spacetime dimensions in a special radial direction of a throat amongst the six extra dimensions of string theory while $f(\phi)$ is known as the warp factor of the throat. Here however, we simply regard \refeq{P-DBI} as a phenomenological model.

The most salient feature of the Lagrangian \refeq{P-DBI} is that its non canonical structure imposes a speed limit on the inflaton $\dot \phi^2 < 1/f(\phi)$. Working out the derivatives in \refeq{cs}, one actually finds that the speed of sound $\cs$ is given by
\be
c_s=\sqrt{1-f(\phi) \dot \phi^2}\,
\ee
so that it is both positive and less than unity. Using the Hamilton-Jacobi equations \cite{DBI2}
\bea
\dot \phi&=&-2 \cs(\phi) \frac{d  H(\phi)}{d \phi} \\
3 H^2(\phi)&=&\frac{1}{f(\phi)}\left(\frac{1}{c_s(\phi)}-1 \right)+V(\phi) \, \qquad {\rm where} \\
c_s(\phi)&=&\left( 1+ 4 f(\phi)  \frac{d  H(\phi)}{d \phi}  \right)^{-1/2}
\eea
one can easily show that requiring $\eps$ ($< 1$ to realize inflation) and $\cs$ to be constant imposes the following specific form for the warp factor and potential
\bea
f(\phi)&=&\frac{1}{2 H_0^2 \epsilon \cs}(1-\cs^2)\, {\rm exp} \left( \mp \sqrt{\frac{2 \epsilon}{\cs}} \phi\right) \\
V(\phi)&=&3 H_0^2 \left(1-\frac{2 \epsilon}{3(1+\cs)} \right){\rm exp} \left(  \pm \sqrt{\frac{2 \epsilon}{\cs}} \phi\right)
\eea
as well as the following background evolution
\be
\dot \phi= \mp H_0 \, \sqrt{2 \epsilon \cs} \, {\rm exp} \left( \pm \sqrt{\frac{\epsilon}{2\cs}} \phi\right)
\ee

\be
H=H_0 \, {\rm exp} \left( \pm \sqrt{\frac{\epsilon}{2\cs}} \phi\right)\,.
\ee

\section{Useful properties of Hankel functions}
\label{Hankel}

Here, we collect the useful properties of the Hankel functions that we used in the calculations of section \ref{verification}.

\be
H^{(1)}_{3/2}(x)=-i \sqrt{\frac{2}{\pi}}e^{i x} \frac{1-i x}{x^{3/2}}
\ee

\be
H_{\nu}^{(2)}(x)=H_{\nu}^{(1)*}(x)
\ee

\be
x \frac{\rm d}{{\rm d} x} H_{\nu}^{(1)}(x)+\nu H_{\nu}^{(1)}(x)=x H_{\nu-1}^{(1)}(x)\,,
\label{deriv}
\ee

\be
\int dx \, x\,  \left( H_{\nu}^{(2)}(x) \right)^2=\frac{x^2}{2} \left[  \left( H_{\nu}^{(2)}(x) \right)^2 -H_{\nu-1}^{(1)}(x) H_{\nu+1}^{(1)}(x) \right]
\label{integr}
\ee

\be
H_{\nu}^{(1)}(x) \to \frac{1}{\Gamma(\nu+1)}\left( \frac{x}{2} \right)^{\nu}-i \frac{\Gamma(\nu)}{\pi} \left( \frac{2}{x} \right)^{\nu} \, \qquad {\rm small } \, x
\label{Hankel-0}
\ee

\be
H_{\nu}^{(1)}(x) \to \sqrt{ \frac{2}{\pi z}} e^{i(z-\frac{\pi}{2}\nu-\frac{\pi}{4})}  \, \qquad {\rm large \,  | z|}\, ,  |{\rm arg\, z} | < \pi
\ee

\section{Useful integrals}
\label{integrals}

\be
 -i \int_{-\infty}^0 d \t  e^{i \t} =-1
\ee

\be
 \int_{-\infty}^0 d \t \t  e^{i \t} =1
\ee

\be
 -i \int_{-\infty}^0 d \t  e^{i \t}  \ln(-\t) =\gamma
\ee

\be
 \int_{-\infty}^0 d \t  e^{i \t}  \t \ln(-\t) =1-\gamma
\ee

\be
{\rm Re} \left[ -i \int_{-\infty}^0 \frac{d \t}{\t^2}(1-i \t)e^{i \t} \right]=-1
\ee

\be
{\rm Re} \left[ -i \int_{-\infty}^0 \frac{d \t}{\t^2}(1-i \t) \ln(-\t)e^{i \t} \right]=-1+\gamma
\ee

\section{Details on the correction terms}
\label{correction-details}

In this appendix, we give the details of the correction terms that are needed in subsection \ref{general-first-order} to calculate $F$ \refeq{F} up to order $\CO(\eps)$. 

We begin by considering the contribution to $\F_1$ coming from the correction to the leading-order scale factor $a \approx -\frac{1}{H \t}$. Plugging the correction terms of \refeq{corr-a} into \refeq{F1} and evaluating $g_1$ and the mode functions to leading order, one finds

\begin{itemize}

\item $\Delta_a \F_1$: 
\bea
\frac{\Delta_a \F_1}{\Pz(\k)}&=&6 \e \invc \k \cs {\rm Re} \left[ -i \int_{-\infty}^0 d \t (1-\ln(-\a k \c \t))e^{2i k c_s \t}\right]    \nn \\
&=&-3 \e \invc (1+\gamma-\l)\,.
\label{DaF1}
\eea

\end{itemize}

Similarly, plugging the correction to $g_1$ (\textit{c.f.} \refeq{corr-g})
 \bea
\Delta g_1(\t)= \left(-3 \frac{\e}{\c^2} \left(\frac{1}{\c^2}-1\right)(\et-4s)  +6 \frac{\e s}{\c^2}   \right)  \ln \frac{\t}{\t_K}
\eea
 into \refeq{F1} gives
 \begin{itemize}
 
\item $\Delta_g \F_1$:

\bea
\frac{\Delta_g \F_1}{\Pz(\k)}&=&\left[- 3(\et-4 s) \left(\frac{1}{\c^2}-1\right)+6 s\right]  \k \c {\rm Re} \left[ -i \int_{-\infty}^0 d \t \ln(-\a k c_{s} \t)e^{2i k c_s \t}\right]  \nn \\
&=&\invc \left(\gamma-\l\right)\left(-\frac{3\eta}{2}+6s\right)+3s\left(\gamma-\l\right)
\label{DgF1}
\eea

\end{itemize}

As for the corrections to the derivative of the mode function \refeq{Du*}, this contributes
\begin{itemize}

\item $\Delta_u \F_1$:

\bea
\frac{\Delta_u \F_1}{\Pz(\k)}&=&6 \left(\frac{1}{\c^2}-1\right) {\rm Re} \left[  \int_{-\infty}^0 \frac{d \t}{\t} e^{2i k c_s \t} \times \left[ (\epsilon +\frac{\eta}{2} +\frac{s}{2})(\frac{i}{x}-1) - i\epsilon x + s x^2  
 \right. \right.
\cr
&&
\left. \left.
+ \left( i(\epsilon +\frac{\eta}{2} - \frac{3s}{2}) - s x\right) x
\ln\frac{\tau}{\tau_K} 
+(\epsilon +\frac{\eta}{2} +\frac{s}{2} )e^{ix} 
\frac{d h^*(x)}{dx} \right]   \right]     \nn \\
&=&\frac32 \invc \left[  \eta \left(\gamma-\l \right)+2 \e\left(1+\gamma-\l \right)-2s\left(\gamma-\l \right) \right] \nn \\
&+& \left(\frac{1}{\c^2}-1\right) (2\e+\eta+s) {\rm Re} \left[   \int_{0}^{+\infty} \frac{dx}{x^2}e^{-2 ix}(x-i)-\int_{0}^{+\infty} \frac{dx}{x}e^{-ix}\frac{d}{dx} h^*(x)   \right] 
\label{DuF1}\nn
\eea
where we have isolated the contributions coming from the first and the last term in brackets in the right-hand side of the first equality. Using the asymptotic behaviour
\be
h^*(x) \to (2-\gamma)i-i  \ln (2x)+\CO(x^2)\,,
\label{h-0}
\ee
one indeed sees that the divergences of the two integrals in the last line of \refeq{DuF1} cancel each other. Using
\bea
\frac{dh^*(x)}{dx} = (1-\frac{\pi}{2} x - \frac{i}{x}) e^{-ix}
- ix e^{ix} {\rm Ei}(-2ix) ~.
\eea
and making use of the integral
\bea
-i \int_0^\infty dx ~{\rm Ei}(-2ix) =\half \,,
\label{Ei-int}
\eea
one evaluates 
\be
{\rm Re} \left[   \int_{0}^{+\infty} \frac{dx}{x^2}e^{-2 ix}(x-i)-\int_{0}^{+\infty} \frac{dx}{x}e^{-ix}\frac{d}{dx} h^*(x)   \right] =-\frac12
\ee
so that
\be
\frac{\Delta_u \F_1}{\Pz(\k)}=\frac32 \invc \left[  \eta \left(-1+\gamma-\l \right)+2 \e \left(\gamma-\l \right)-s \left(1+2\left(\gamma-\l \right) \right) \right] \,.
\label{DuF1}
\ee

\end{itemize}

The same procedure for the $\F_2$ term \refeq{F2} gives

\begin{itemize}

\item $\Delta_a \F_2$:

\bea
\frac{\Delta_a \F_2}{\Pz(\k)}&=&2 \e \invc \frac{1}{\k \cs} {\rm Re} \left[ -i \int_{-\infty}^0  \frac{d \t}{\t^2} (-1+\ln(-\a k c_{s} \t))  (1-i k c_s \t)^2 e^{2i k c_s \t}\right]    \nn \\
&=&-\e \invc \left(1-3\left(\gamma-\l \right) \right)
\label{DaF2}
\eea

\item $\Delta_g \F_2$:

\bea
\Delta g_2(\t)=  \left[ \e \left(\frac{1}{\c^2}-1\right)(\et-2s)  -2 \e s \right] \ln \frac{\t}{\t_K}  
\eea

\bea
\frac{\Delta_g \F_2}{\Pz(\k)}&=& \left[ (\et-2 s) \left(\frac{1}{\c^2}-1\right)-2s \right]  \frac{1}{ \c \k}   {\rm Re} \left[ -i \int_{-\infty}^0 \frac{d \t}{\t^2} \ln(-\a k c_{s} \t)(1-i \k \c \t)^2e^{2i k c_s \t}\right]      \nn \\
&=&  \left[ (\et-2 s) \left(\frac{1}{\c^2}-1\right)-2s \right] \left(-2+\frac32\left(\gamma-\l\right)   \right)
\label{DgF2}
\eea

\item $\Delta_u \F_2$:

\bea
\frac{\Delta_u \F_2}{\Pz(\k)}&=&2 \left(\frac{1}{\c^2}-1\right) \frac{1}{k \c} {\rm Re} \left[  \int_{-\infty}^0 \frac{d \t}{\t^2}(1-i \k \c \t) e^{2i k \c \t} 
\right.
\cr
&\times &
\left. 
 \left[(\epsilon + s) (x-i) + i s x^2 +\left(-(\epsilon +\frac{\eta}{2} +\frac{s}{2})(x-i)  - i x^2 s\right) \ln\frac{\tau}{\tau_K}
\right. \right.
\cr
&&
\left. \left.
+   (\epsilon +\frac{\eta}{2}
+\frac{s}{2})e^{ix} h^*(x) \right]   \right]   \nn \\
&=&\frac12 \invc \left[  \eps \left(2-6\left(\gamma-\l \right) \right)+ \eta\left( 4 -3\left(\gamma-\l \right) \right)\right]\nn \\
&+& \left(\frac{1}{\c^2}-1\right) (2\e+\eta+s) {\rm Re} \left[   \int_{0}^{+\infty} \frac{dx}{x^2}e^{-ix}(1+ix) h^*(x)   \right] 
\eea
where it is clear from the asymptotic behaviour \refeq{h-0} that the last integral is convergent. Using \refeq{Ei-int},
\bea
h^*(x) = 2ie^{-ix} - \frac{\pi}{2} (1+ix)e^{-ix} 
-ie^{ix} (1-ix) {\rm Ei}(-2ix)  
\eea
and
\bea
\int_0^\infty \frac{dx}{x^2} \left[ (-\frac{\pi}{2}-2x) \cos 2x 
+ (2-\pi x) \sin 2x - {\rm si}(2x) \right] = 0 \,,
\eea
one finds
\be
{\rm Re} \left[   \int_{0}^{+\infty} \frac{dx}{x^2}e^{-ix}(1+ix) h^*(x)   \right] =\frac12
\ee
so that eventually
\be
\frac{\Delta_u \F_2}{\Pz(\k)}=\frac12 \invc \left[2 \e \left( 2 -3\left(\gamma-\l \right) \right)+  \eta \left(5-3\left(\gamma-\l \right) \right)+s \right]
\label{DuF2}
\ee

\end{itemize}

Summing the six contributions above, one finds that the terms in $\l$ disappear, as expected, and one gets
\be
\frac{\Delta \F}{\Pz(k)}=-\invc (2 \e+\eta-3s)+4s \,.
\ee

\bibliography{Biblio}

\begin{thebibliography}{34}
\expandafter\ifx\csname natexlab\endcsname\relax\def\natexlab#1{#1}\fi
\expandafter\ifx\csname bibnamefont\endcsname\relax
  \def\bibnamefont#1{#1}\fi
\expandafter\ifx\csname bibfnamefont\endcsname\relax
  \def\bibfnamefont#1{#1}\fi
\expandafter\ifx\csname citenamefont\endcsname\relax
  \def\citenamefont#1{#1}\fi
\expandafter\ifx\csname url\endcsname\relax
  \def\url#1{\texttt{#1}}\fi
\expandafter\ifx\csname urlprefix\endcsname\relax\def\urlprefix{URL }\fi
\providecommand{\bibinfo}[2]{#2}
\providecommand{\eprint}[2][]{\url{#2}}

\bibitem[{\citenamefont{Komatsu et~al.}(2010)}]{Komatsu:2010fb}
\bibinfo{author}{\bibfnamefont{E.}~\bibnamefont{Komatsu}} \bibnamefont{et~al.}
  (\bibinfo{year}{2010}), \eprint{1001.4538}.

\bibitem[{\citenamefont{Chen}(2010)}]{Chen:2010xk}
\bibinfo{author}{\bibfnamefont{X.}~\bibnamefont{Chen}} (\bibinfo{year}{2010}),
  \eprint{1002.1416}.

\bibitem[{\citenamefont{Koyama}(2010)}]{Koyama:2010xj}
\bibinfo{author}{\bibfnamefont{K.}~\bibnamefont{Koyama}}
  (\bibinfo{year}{2010}), \eprint{1002.0600}.

\bibitem[{\citenamefont{Liguori et~al.}(2010)\citenamefont{Liguori, Sefusatti,
  Fergusson, and Shellard}}]{Liguori:2010hx}
\bibinfo{author}{\bibfnamefont{M.}~\bibnamefont{Liguori}},
  \bibinfo{author}{\bibfnamefont{E.}~\bibnamefont{Sefusatti}},
  \bibinfo{author}{\bibfnamefont{J.~R.} \bibnamefont{Fergusson}},
  \bibnamefont{and} \bibinfo{author}{\bibfnamefont{E.~P.~S.}
  \bibnamefont{Shellard}} (\bibinfo{year}{2010}), \eprint{1001.4707}.

\bibitem[{\citenamefont{Komatsu}(2010)}]{Komatsu:2010hc}
\bibinfo{author}{\bibfnamefont{E.}~\bibnamefont{Komatsu}}
  (\bibinfo{year}{2010}), \eprint{1003.6097}.

\bibitem[{\citenamefont{Yadav and Wandelt}(2010)}]{Yadav:2010fz}
\bibinfo{author}{\bibfnamefont{A.~P.~S.} \bibnamefont{Yadav}} \bibnamefont{and}
  \bibinfo{author}{\bibfnamefont{B.~D.} \bibnamefont{Wandelt}}
  (\bibinfo{year}{2010}), \eprint{1006.0275}.

\bibitem[{\citenamefont{Maldacena}(2003)}]{malda}
\bibinfo{author}{\bibfnamefont{J.~M.} \bibnamefont{Maldacena}},
  \bibinfo{journal}{JHEP} \textbf{\bibinfo{volume}{05}}, \bibinfo{pages}{013}
  (\bibinfo{year}{2003}), \eprint{astro-ph/0210603}.

\bibitem[{\citenamefont{Creminelli and Zaldarriaga}(2004)}]{Creminelli:2004yq}
\bibinfo{author}{\bibfnamefont{P.}~\bibnamefont{Creminelli}} \bibnamefont{and}
  \bibinfo{author}{\bibfnamefont{M.}~\bibnamefont{Zaldarriaga}},
  \bibinfo{journal}{JCAP} \textbf{\bibinfo{volume}{0410}}, \bibinfo{pages}{006}
  (\bibinfo{year}{2004}), \eprint{astro-ph/0407059}.

\bibitem[{\citenamefont{Seery and Lidsey}(2005)}]{Seery:2005wm}
\bibinfo{author}{\bibfnamefont{D.}~\bibnamefont{Seery}} \bibnamefont{and}
  \bibinfo{author}{\bibfnamefont{J.~E.} \bibnamefont{Lidsey}},
  \bibinfo{journal}{JCAP} \textbf{\bibinfo{volume}{0506}}, \bibinfo{pages}{003}
  (\bibinfo{year}{2005}), \eprint{astro-ph/0503692}.

\bibitem[{\citenamefont{Chen et~al.}(2007{\natexlab{a}})\citenamefont{Chen,
  Huang, Kachru, and Shiu}}]{Chen:2006nt}
\bibinfo{author}{\bibfnamefont{X.}~\bibnamefont{Chen}},
  \bibinfo{author}{\bibfnamefont{M.-x.} \bibnamefont{Huang}},
  \bibinfo{author}{\bibfnamefont{S.}~\bibnamefont{Kachru}}, \bibnamefont{and}
  \bibinfo{author}{\bibfnamefont{G.}~\bibnamefont{Shiu}},
  \bibinfo{journal}{JCAP} \textbf{\bibinfo{volume}{0701}}, \bibinfo{pages}{002}
  (\bibinfo{year}{2007}{\natexlab{a}}), \eprint{hep-th/0605045}.

\bibitem[{\citenamefont{Cheung et~al.}(2008{\natexlab{a}})\citenamefont{Cheung,
  Fitzpatrick, Kaplan, and Senatore}}]{Cheung:2007sv}
\bibinfo{author}{\bibfnamefont{C.}~\bibnamefont{Cheung}},
  \bibinfo{author}{\bibfnamefont{A.~L.} \bibnamefont{Fitzpatrick}},
  \bibinfo{author}{\bibfnamefont{J.}~\bibnamefont{Kaplan}}, \bibnamefont{and}
  \bibinfo{author}{\bibfnamefont{L.}~\bibnamefont{Senatore}},
  \bibinfo{journal}{JCAP} \textbf{\bibinfo{volume}{0802}}, \bibinfo{pages}{021}
  (\bibinfo{year}{2008}{\natexlab{a}}), \eprint{0709.0295}.

\bibitem[{\citenamefont{Ganc and Komatsu}(2010)}]{Ganc:2010ff}
\bibinfo{author}{\bibfnamefont{J.}~\bibnamefont{Ganc}} \bibnamefont{and}
  \bibinfo{author}{\bibfnamefont{E.}~\bibnamefont{Komatsu}}
  (\bibinfo{year}{2010}), \eprint{1006.5457}.

\bibitem[{\citenamefont{Garriga and Mukhanov}(1999)}]{Garriga:1999vw}
\bibinfo{author}{\bibfnamefont{J.}~\bibnamefont{Garriga}} \bibnamefont{and}
  \bibinfo{author}{\bibfnamefont{V.~F.} \bibnamefont{Mukhanov}},
  \bibinfo{journal}{Phys. Lett.} \textbf{\bibinfo{volume}{B458}},
  \bibinfo{pages}{219} (\bibinfo{year}{1999}), \eprint{hep-th/9904176}.

\bibitem[{\citenamefont{Langlois and Renaux-Petel}(2008)}]{Langlois2008}
\bibinfo{author}{\bibfnamefont{D.}~\bibnamefont{Langlois}} \bibnamefont{and}
  \bibinfo{author}{\bibfnamefont{S.}~\bibnamefont{Renaux-Petel}},
  \bibinfo{journal}{JCAP} \textbf{\bibinfo{volume}{0804}}, \bibinfo{pages}{017}
  (\bibinfo{year}{2008}), \eprint{0801.1085}.

\bibitem[{\citenamefont{Gao}(2008)}]{Gao:2008dt}
\bibinfo{author}{\bibfnamefont{X.}~\bibnamefont{Gao}}, \bibinfo{journal}{JCAP}
  \textbf{\bibinfo{volume}{0806}}, \bibinfo{pages}{029} (\bibinfo{year}{2008}),
  \eprint{0804.1055}.

\bibitem[{\citenamefont{Langlois et~al.}(2008)\citenamefont{Langlois,
  Renaux-Petel, Steer, and Tanaka}}]{Langlois:2008qf}
\bibinfo{author}{\bibfnamefont{D.}~\bibnamefont{Langlois}},
  \bibinfo{author}{\bibfnamefont{S.}~\bibnamefont{Renaux-Petel}},
  \bibinfo{author}{\bibfnamefont{D.~A.} \bibnamefont{Steer}}, \bibnamefont{and}
  \bibinfo{author}{\bibfnamefont{T.}~\bibnamefont{Tanaka}},
  \bibinfo{journal}{Phys. Rev.} \textbf{\bibinfo{volume}{D78}},
  \bibinfo{pages}{063523} (\bibinfo{year}{2008}), \eprint{0806.0336}.

\bibitem[{\citenamefont{Arroja et~al.}(2008)\citenamefont{Arroja, Mizuno, and
  Koyama}}]{Arroja:2008yy}
\bibinfo{author}{\bibfnamefont{F.}~\bibnamefont{Arroja}},
  \bibinfo{author}{\bibfnamefont{S.}~\bibnamefont{Mizuno}}, \bibnamefont{and}
  \bibinfo{author}{\bibfnamefont{K.}~\bibnamefont{Koyama}},
  \bibinfo{journal}{JCAP} \textbf{\bibinfo{volume}{0808}}, \bibinfo{pages}{015}
  (\bibinfo{year}{2008}), \eprint{0806.0619}.

\bibitem[{\citenamefont{Arnowitt et~al.}(1962)\citenamefont{Arnowitt, Deser,
  and Misner}}]{adm}
\bibinfo{author}{\bibfnamefont{R.}~\bibnamefont{Arnowitt}},
  \bibinfo{author}{\bibfnamefont{S.}~\bibnamefont{Deser}}, \bibnamefont{and}
  \bibinfo{author}{\bibfnamefont{C.~W.} \bibnamefont{Misner}},
  \bibinfo{journal}{Louis Witten ed} \textbf{\bibinfo{volume}{7}},
  \bibinfo{pages}{227} (\bibinfo{year}{1962}).

\bibitem[{\citenamefont{Chen et~al.}(2007{\natexlab{b}})\citenamefont{Chen,
  Easther, and Lim}}]{Chen:2006xjb}
\bibinfo{author}{\bibfnamefont{X.}~\bibnamefont{Chen}},
  \bibinfo{author}{\bibfnamefont{R.}~\bibnamefont{Easther}}, \bibnamefont{and}
  \bibinfo{author}{\bibfnamefont{E.~A.} \bibnamefont{Lim}},
  \bibinfo{journal}{JCAP} \textbf{\bibinfo{volume}{0706}}, \bibinfo{pages}{023}
  (\bibinfo{year}{2007}{\natexlab{b}}), \eprint{astro-ph/0611645}.

\bibitem[{\citenamefont{Chen et~al.}(2008)\citenamefont{Chen, Easther, and
  Lim}}]{Chen:2008wn}
\bibinfo{author}{\bibfnamefont{X.}~\bibnamefont{Chen}},
  \bibinfo{author}{\bibfnamefont{R.}~\bibnamefont{Easther}}, \bibnamefont{and}
  \bibinfo{author}{\bibfnamefont{E.~A.} \bibnamefont{Lim}},
  \bibinfo{journal}{JCAP} \textbf{\bibinfo{volume}{0804}}, \bibinfo{pages}{010}
  (\bibinfo{year}{2008}), \eprint{0801.3295}.

\bibitem[{\citenamefont{Mukhanov et~al.}(1992)\citenamefont{Mukhanov, Feldman,
  and Brandenberger}}]{MFB}
\bibinfo{author}{\bibfnamefont{V.~F.} \bibnamefont{Mukhanov}},
  \bibinfo{author}{\bibfnamefont{H.~A.} \bibnamefont{Feldman}},
  \bibnamefont{and} \bibinfo{author}{\bibfnamefont{R.~H.}
  \bibnamefont{Brandenberger}}, \bibinfo{journal}{Phys. Rept.}
  \textbf{\bibinfo{volume}{215}}, \bibinfo{pages}{203} (\bibinfo{year}{1992}).

\bibitem[{\citenamefont{Langlois}(2010)}]{Langlois:2010xc}
\bibinfo{author}{\bibfnamefont{D.}~\bibnamefont{Langlois}}
  (\bibinfo{year}{2010}), \eprint{1001.5259}.

\bibitem[{\citenamefont{Keldysh}(1964)}]{Keldysh:1964ud}
\bibinfo{author}{\bibfnamefont{L.~V.} \bibnamefont{Keldysh}},
  \bibinfo{journal}{Zh. Eksp. Teor. Fiz.} \textbf{\bibinfo{volume}{47}},
  \bibinfo{pages}{1515} (\bibinfo{year}{1964}).

\bibitem[{\citenamefont{Schwinger}(1961)}]{Schwinger:1960qe}
\bibinfo{author}{\bibfnamefont{J.~S.} \bibnamefont{Schwinger}},
  \bibinfo{journal}{J. Math. Phys.} \textbf{\bibinfo{volume}{2}},
  \bibinfo{pages}{407} (\bibinfo{year}{1961}).

\bibitem[{\citenamefont{Weinberg}(2005)}]{Weinberg:2005vy}
\bibinfo{author}{\bibfnamefont{S.}~\bibnamefont{Weinberg}},
  \bibinfo{journal}{Phys. Rev.} \textbf{\bibinfo{volume}{D72}},
  \bibinfo{pages}{043514} (\bibinfo{year}{2005}), \eprint{hep-th/0506236}.

\bibitem[{\citenamefont{Spalinski}(2008)}]{Spalinski:2007un}
\bibinfo{author}{\bibfnamefont{M.}~\bibnamefont{Spalinski}},
  \bibinfo{journal}{JCAP} \textbf{\bibinfo{volume}{0804}}, \bibinfo{pages}{002}
  (\bibinfo{year}{2008}), \eprint{0711.4326}.

\bibitem[{\citenamefont{Kinney and Tzirakis}(2008)}]{Kinney:2007ag}
\bibinfo{author}{\bibfnamefont{W.~H.} \bibnamefont{Kinney}} \bibnamefont{and}
  \bibinfo{author}{\bibfnamefont{K.}~\bibnamefont{Tzirakis}},
  \bibinfo{journal}{Phys. Rev.} \textbf{\bibinfo{volume}{D77}},
  \bibinfo{pages}{103517} (\bibinfo{year}{2008}), \eprint{0712.2043}.

\bibitem[{\citenamefont{Cheung et~al.}(2008{\natexlab{b}})\citenamefont{Cheung,
  Creminelli, Fitzpatrick, Kaplan, and Senatore}}]{Cheung:2007st}
\bibinfo{author}{\bibfnamefont{C.}~\bibnamefont{Cheung}},
  \bibinfo{author}{\bibfnamefont{P.}~\bibnamefont{Creminelli}},
  \bibinfo{author}{\bibfnamefont{A.~L.} \bibnamefont{Fitzpatrick}},
  \bibinfo{author}{\bibfnamefont{J.}~\bibnamefont{Kaplan}}, \bibnamefont{and}
  \bibinfo{author}{\bibfnamefont{L.}~\bibnamefont{Senatore}},
  \bibinfo{journal}{JHEP} \textbf{\bibinfo{volume}{03}}, \bibinfo{pages}{014}
  (\bibinfo{year}{2008}{\natexlab{b}}), \eprint{0709.0293}.

\bibitem[{\citenamefont{Silverstein and Tong}(2004)}]{DBI1}
\bibinfo{author}{\bibfnamefont{E.}~\bibnamefont{Silverstein}} \bibnamefont{and}
  \bibinfo{author}{\bibfnamefont{D.}~\bibnamefont{Tong}},
  \bibinfo{journal}{Phys. Rev.} \textbf{\bibinfo{volume}{D70}},
  \bibinfo{pages}{103505} (\bibinfo{year}{2004}), \eprint{hep-th/0310221}.

\bibitem[{\citenamefont{Alishahiha et~al.}(2004)\citenamefont{Alishahiha,
  Silverstein, and Tong}}]{DBI2}
\bibinfo{author}{\bibfnamefont{M.}~\bibnamefont{Alishahiha}},
  \bibinfo{author}{\bibfnamefont{E.}~\bibnamefont{Silverstein}},
  \bibnamefont{and} \bibinfo{author}{\bibfnamefont{D.}~\bibnamefont{Tong}},
  \bibinfo{journal}{Phys. Rev.} \textbf{\bibinfo{volume}{D70}},
  \bibinfo{pages}{123505} (\bibinfo{year}{2004}), \eprint{hep-th/0404084}.

\bibitem[{\citenamefont{Wands}(2010)}]{Wands:2010af}
\bibinfo{author}{\bibfnamefont{D.}~\bibnamefont{Wands}} (\bibinfo{year}{2010}),
  \eprint{1004.0818}.

\bibitem[{\citenamefont{Byrnes and Choi}(2010)}]{Byrnes:2010em}
\bibinfo{author}{\bibfnamefont{C.~T.} \bibnamefont{Byrnes}} \bibnamefont{and}
  \bibinfo{author}{\bibfnamefont{K.-Y.} \bibnamefont{Choi}}
  (\bibinfo{year}{2010}), \eprint{1002.3110}.

\bibitem[{\citenamefont{Lehners}(2010)}]{Lehners:2010fy}
\bibinfo{author}{\bibfnamefont{J.-L.} \bibnamefont{Lehners}}
  (\bibinfo{year}{2010}), \eprint{1001.3125}.

\bibitem[{\citenamefont{Bernardeau}(2010)}]{Bernardeau:2010jp}
\bibinfo{author}{\bibfnamefont{F.}~\bibnamefont{Bernardeau}}
  (\bibinfo{year}{2010}), \eprint{1003.2869}.

\end{thebibliography}

\end{document}